\documentclass[print,aps,prd,superscriptaddress,nofootinbib]{revtex4-2}

\usepackage[utf8]{inputenc}
\usepackage[english]{babel}

\usepackage{amsmath,amssymb,mathtools, dcolumn, slashed, graphicx, braket, bm, xcolor, url, hyperref}
 
\definecolor{persianindigo}{rgb}{0.2, 0.07, 0.48}
\definecolor{plum(traditional)}{rgb}{0.56, 0.27, 0.52}
\definecolor{purplemountainmajesty}{rgb}{0.59, 0.47, 0.71}
\definecolor{raspberryrose}{rgb}{0.7, 0.27, 0.42}
\definecolor{ruby}{rgb}{0.88, 0.07, 0.37}
\definecolor{bluebell}{rgb}{0.64, 0.64, 0.82}
\definecolor{ballblue}{rgb}{0.13, 0.67, 0.8}
\definecolor{blue(ncs)}{rgb}{0.0, 0.53, 0.74}
\definecolor{blue(pigment)}{rgb}{0.2, 0.2, 0.6}
\definecolor{forestgreen(web)}{rgb}{0.13, 0.55, 0.13}

\definecolor{pastelpink}{rgb}{0.75, 0.15, 0.49}

\hypersetup{colorlinks=true,		
	linkcolor=blue(pigment),	          				
	citecolor=blue(pigment),					
	filecolor=blue(pigment),							
	urlcolor=blue(pigment)}							

\def\d{{\rm d}}

\def\xB{x_{\scriptscriptstyle B}}

\def\i-{\item[-]}

\def\g{\gamma}

\def\l{\lambda}

\def\Re{{\rm Re}}

\def\1{\mathbb 1}

\usepackage{float}
\usepackage[normalem]{ulem}
\definecolor{luca_comm}{RGB}{199, 104, 10}
\definecolor{umb_comm}{RGB}{60, 176, 78}
\definecolor{raj_comm}{RGB}{191, 0, 0}
\definecolor{franc_comm}{RGB}{150, 15, 15}
\definecolor{cris_comm}{RGB}{0, 5, 75}

\newcommand{\myparallel}{{\mkern3mu\vphantom{\perp}\vrule depth 0pt\mkern2mu\vrule depth 0pt\mkern3mu}}

\begin{document}

\title{$J/\psi$ polarization in large-$P_T$ semi-inclusive deep-inelastic scattering at the EIC}

\author{Umberto D'Alesio}
\email{umberto.dalesio@ca.infn.it}
\affiliation{Dipartimento di Fisica, Universit\`a di Cagliari, Cittadella Universitaria, I-09042 Monserrato (CA), Italy}
\affiliation{INFN, Sezione di Cagliari, Cittadella Universitaria, I-09042 Monserrato (CA), Italy}

\author{Luca Maxia}
\email{luca.maxia@ca.infn.it}
\affiliation{Dipartimento di Fisica, Universit\`a di Cagliari, Cittadella Universitaria, I-09042 Monserrato (CA), Italy}
\affiliation{INFN, Sezione di Cagliari, Cittadella Universitaria, I-09042 Monserrato (CA), Italy}

\author{Francesco Murgia}
\email{francesco.murgia@ca.infn.it}
\affiliation{INFN, Sezione di Cagliari, Cittadella Universitaria, I-09042 Monserrato (CA), Italy}

\author{Cristian Pisano}
\email{cristian.pisano@unica.it}
\affiliation{Dipartimento di Fisica, Universit\`a di Cagliari, Cittadella Universitaria, I-09042 Monserrato (CA), Italy}
\affiliation{INFN, Sezione di Cagliari, Cittadella Universitaria, I-09042 Monserrato (CA), Italy}

\author{Sangem Rajesh}
\email{sangem.rajesh@vit.ac.in}
 \affiliation{Department of Physics, School of Advanced Sciences, Vellore Institute of Technology, Vellore,
Tamil Nadu 632014, India}
\affiliation{INFN, Sezione di Perugia, via A.~Pascoli snc, 06123, Perugia, Italy}

\begin{abstract}
We present a detailed phenomenological study of $J/\psi$ polarization in semi-inclusive deep inelastic scattering processes, focusing on the kinematics accessible at the future Electron-Ion Collider. We show theoretical estimates for the standard polarization parameters for different frames usually adopted in the literature, in the large $P_T$ region, namely $P_T\gg \Lambda_\text{QCD}$, where collinear factorization is expected to hold.
We adopt both the Color Singlet Model and the Nonrelativistic QCD approach, paying special attention to the role of different sets of Long Distance Matrix Elements. Finally we present a preliminary analysis of some frame independent polarization invariants.
\end{abstract}

\date{\today}
\maketitle

\section{Introduction}

Our understanding of the $J/\psi$ production mechanism at high energies has improved significantly since its discovery almost 50 years ago~\cite{E598:1974sol,SLAC-SP-017:1974ind}, thanks to the combined efforts from both the theoretical and experimental communities. However, there are still major problems in the theoretical analyses of the available data, such as the long-standing $J/\psi$ polarization puzzle. Namely, $J/\psi$ polarization measurements cannot yet be explained in a way entirely consistent with the world experimental results for the unpolarized $J/\psi$ yields.

The present theoretical frameworks all agree in providing a perturbative description of  the creation of the charm quark-antiquark ($c \bar c$) pair. The charm mass $m_c$ plays the role of the hard scale, since it is much larger than the asymptotic scale parameter of QCD, $\Lambda_\text{QCD}$. These approaches nonetheless differ in the treatment of the subsequent nonperturbative  transition to the hadronic bound state. For instance, in the traditional Color-Singlet Model (CSM)~\cite{Baier:1983va} the $c \bar c$ pair is produced at short distances directly with the quantum numbers of the $J/\psi$ meson, {\it i.e.}\ in a color-singlet (CS) state with spin one and no orbital angular momentum. This is possible by the emission of an additional hard gluon, which implies the suppression of the cross section by one power of the strong coupling constant $\alpha_s$. However, the CSM cannot be considered as a complete theory, since at the next-to-leading order (NLO) $P$-wave quarkonia are affected by uncanceled infrared singularities.

These singularities are properly removed in the effective field theory approach of nonrelativistic QCD  (NRQCD), based on a rigorous factorization theorem, which was assumed in the original paper by Bodwin, Braaten, and Lepage~\cite{Bodwin:1994jh}, and later explicitly proven to next-to-next-to-leading order (NNLO)~\cite{Nayak:2006fm}. NRQCD therefore implies a separation of process-dependent short-distance coefficients, to be calculated perturbatively as expansions in $\alpha_s$, from long-distance matrix elements (LDMEs), which are expected to be universal and have to be extracted from experiments. Scaling rules~\cite{Lepage:1992tx} predict each of the LDMEs to scale with a definite power of the relative velocity $v$ of the heavy quark-antiquark pair in the quarkonium rest frame in the limit $v \ll 1$.
Observables are hence evaluated by means of a double expansion in $\alpha_s$ and in $v$, with $\alpha_s \simeq 0.2$ and $v^2 \simeq 0.3$ for charmonium states. An essential feature of this approach is that the $c \bar c$ pair at short distance can be produced in any Fock state
$n = \, ^{2S+1} L_J^{[c]}$ with definite orbital angular momentum $L$, spin $S$, total angular momentum $J$ and color configuration $c
= 1, 8$. NRQCD hence predicts the existence of intermediate color-octet
(CO) states, which subsequently evolve into physical, CS quarkonia by the emission of soft gluons. For $S$-wave quarkonia, the CSM is recovered in the limit $v\to 0$.
In the specific case of $J/\psi$ production, the CSM prediction is based only on the $^3S_1^{[1]}$ CS state, while NRQCD includes the leading relativistic corrections as well,
which at the relative order ${\cal O}(v^4)$ are given by the CO states $^1S_0^{[8]},$ $^3S_1^{[8]},$ and $^3P_J^{[8]}$ with $J = 0,1,2$.

The values of the CO LDMEs extracted from different fits to data on $J/\psi$ and $\Upsilon$ yields~\cite{Butenschoen:2010rq,Chao:2012iv,Sharma:2012dy,Bodwin:2014gia,Zhang:2014ybe} are not compatible with each other, even within the large uncertainties~\cite{Brambilla:2010cs,Andronic:2015wma,Lansberg:2019adr}. Therefore, any new method to determine them with better precision is worth exploring~\cite{Zhang:2019ecf,Qiu:2020xum,Boer:2021ehu}. In this paper we propose to look at the $J/\psi$ polarization parameters in semi-inclusive deep-inelastic scattering (SIDIS), $e\, p \to e'\,J/\psi\, X$, in a kinematic region where  the transverse momentum of the $J/\psi$ meson $P_T$ is large, namely $P_T \gg \Lambda_\text{QCD}$, and collinear factorization is expected to hold. Analysing SIDIS at finite values of the exchanged photon virtuality $Q^2$ has certain experimental and theoretical advantages as compared to photoproduction. Namely, as $Q^2$ increases theoretical uncertainties in the different frameworks decrease and resolved photon contributions are expected to be negligible. Moreover, background from diffractive $J/\psi$ production is expected to decrease with $Q^2$ faster than the SIDIS cross section. The distinct signature of the scattered lepton makes the process particularly easy to detect. Clearly, cross sections are smaller than those expected in the photoproduction case, however, considering the achievable high luminosities, this study should be feasible at the future Electron-Ion Collider (EIC) planned in the United States~\cite{AbdulKhalek:2021gbh,Accardi:2012qut,Boer:2011fh}.

So far, only a single experimental study of $J/\psi$ polarization in SIDIS has been performed, by the H1 Collaboration at HERA~\cite{H1:2002xeb}. Such a measurement is limited to the polarization parameter $\lambda$ in the helicity frame. This result turns out to be compatible with the predictions provided in Refs.~\cite{Fleming:1997fq,Yuan:2000cn}, but it can hardly discriminate among the different models. In analogy with Refs.~\cite{Fleming:1997fq,Yuan:2000cn}, our phenomenological analysis has been carried out at the perturbative order $\alpha_s^2$, which has to be considered as the state of the art for these observables. Higher-order effects have been calculated very recently only for the unpolarized cross section within the CSM~\cite{Sun:2017wxk}. Anyway, we expect these effects (at least in the large $Q^2$ region) to be small for the observables we are investigating, because they are ratios of cross sections. We point out that our estimates include also the polarization parameters $\mu$ and $\nu$, not addressed in Refs.~\cite{Fleming:1997fq,Yuan:2000cn}, which are studied in different reference frames. Furthermore, we perform a preliminary study of rotational invariant combinations of these parameters.

The remainder of the paper is organized as follows. In section~\ref{sec: kin_form} we recall the standard SIDIS variables and collect the expressions of the differential cross section for quarkonium production and its leptonic decay in terms of the helicity structure functions and the polarization parameters.
In section~\ref{sec: ang_distr} we discuss the three polarization parameters $\lambda$, $\mu$, $\nu$, showing their estimates in two reference frames and paying special attention to their energy, $z$ and $P_T$ dependences as well as to the impact of the LDME set adopted. To overcome the intrinsic frame dependence of the polarization parameters, in section~\ref{sec: invariants} we present two classes of the so-called rotational invariant quantities, and show, as a case of study, some results for one of them. Finally in section~\ref{sec: conclusions} we gather our conclusions.

\section{Kinematics and Formalism}
\label{sec: kin_form}
In this section we provide the main analytic expressions needed to carry out the phenomenological analysis. For more details and the complete formalism we refer the reader to Ref.~\cite{DAlesio:2021yws}. We consider the SIDIS process
\begin{equation}
	e(k) + p(P) \to e'(k') + J/\psi(P_\psi) + X (P_X) \, ,
	\label{eq: SIDIS Jpsi}
\end{equation}
with the subsequent $J/\psi$ decay into a lepton pair
\begin{equation}
	J/\psi(P_\psi) \to l^+(l) + l^-(l')\, ,
\end{equation}
where, in brackets, we have shown the four-momenta of each particle.
The $J/\psi$ meson is produced via the partonic subprocess
\begin{equation}
\label{eq:parton}
\g^*(q) + a(p_a) \to c\bar c[n] (P_\psi) + a(p_a')\, ,
\end{equation}
with $q^2 =-Q^2$ and $P_\psi^2 = M^2_\psi = (2 m_c)^2$. The initial parton momentum, $p_a$, is related to the parent proton one, $P$, as
\begin{equation}
    p_a = \xi P\, .
\end{equation}
We adopt the following three standard invariant quantities, defined in terms of the photon and hadron momenta
\begin{equation}
\xB = {Q^2 \over 2 P\cdot q}, \quad y = {P \cdot q \over P \cdot k}, \quad z = {P \cdot P_\psi \over P \cdot q},
\end{equation}
where $\xB$ is the Bjorken variable, $y$ is the inelasticity and $z$ is the energy fraction carried out by the $J/\psi$ (in the proton rest frame). All these variables are constrained in the region $ 0 \leq \xB, y, z \leq 1$ and they are connected to other kinematical quantities of the system, like the total center-of-mass (cm) energy $\sqrt s$ and the virtual photon-proton cm energy, $W$.

The cross section that describes the $J/\psi$ formation and its decay into a lepton pair can be written as
\begin{equation}
	{1 \over B_{ll}}{\d \sigma \over \d x_B \,\d y\, \d z\, \d^2 \bm P_T\, \d \Omega} = \frac{\alpha}{ 8\,y \,z \,Q^2} \, \frac{3}{8 \pi} \,
	\left[ {\cal W}_T (1 + \cos^2 \theta) + {\cal W}_L (1 - \cos^2 \theta) + {\cal W}_\Delta \sin 2\theta \cos \phi  + {\cal W}_{\Delta\Delta} \sin^2 \theta \cos 2\phi \right] \,,
	\label{eq: polarization parameterization 1}
\end{equation}
where
$\bm P_T $ is the $J/\psi$ transverse momentum in the cm frame of the virtual photon and the proton, $B_{ll}$ is the branching
ratio for the decay process $J/\psi\to \ell^+\ell^-$ and $\Omega(\theta,\phi)$ refers to the solid angle spanned by the lepton $\ell^+$ in a reference frame where the system formed by $\ell^+$ and $\ell^-$ is at rest.
Moreover, we have introduced the following helicity structure functions
\begin{align}
{\cal W}_T  & \equiv {\cal W}_{1 1} = {\cal W}_{-1,-1}\, , \nonumber \\
{\cal W}_L & \equiv {\cal W}_{0 0}   \, , \nonumber \\
{\cal W}_\Delta & \equiv \frac{1}{\sqrt{2}}\, ({\cal W}_{1 0} + {\cal W}_{0 1}) =  \sqrt{2}\, \text{Re}\, [{\cal W}_{1 0}]\, , \nonumber \\
{\cal W}_{\Delta \Delta} & \equiv {\cal W}_{1, -1} = {\cal W}_{-1, 1} \, ,
\label{eq:W-P-L}
\end{align}
where the subscripts refer to the $J/\psi$ polarization states. More specifically, ${\cal W}_T$ and ${\cal W}_L$ are respectively the structure functions for transversely and longitudinally polarized $J/\psi$ mesons, ${\cal W}_\Delta$ is the single-helicity flip structure function, and ${\cal W}_{\Delta\Delta}$ is the double-helicity flip one. Notice that in Eq.~\eqref{eq: polarization parameterization 1} we have introduced a proper overall constant factor w.r.t.~Eq.~(2.35) of Ref.~\cite{DAlesio:2021yws} to ensure the normalization when integrated over the solid angle, see Eq.~\eqref{eq: unpol xsec} below. This does not affect any conclusion of Ref.~\cite{DAlesio:2021yws}, where all relevant quantities are defined as ratios of helicity structure functions.

As shown in Ref.~\cite{DAlesio:2021yws}, the structure functions in Eq.~\eqref{eq:W-P-L}
can be further decomposed in terms of the contributions coming from the longitudinal ($\myparallel$) and transverse ($\perp$) polarizations of the virtual photon. Moreover, within a collinear factorization scheme, they are given as convolutions of collinear parton distribution functions (PDFs) with partonic helicity structure functions (weighted by proper LDMEs). These, in turn, can be expressed as functions of the partonic Mandelstam invariants.

The unpolarized cross section is obtained by integrating Eq.~\eqref{eq: polarization parameterization 1} over the solid angle $\Omega$,
\begin{equation}
 {1 \over B_{ll}}{\d \sigma \over \d \xB\, \d y\,\d z\,  \d^2 \bm P_T} = {\alpha \over {8 \, y\,  z\,  Q^2}}\, (2{\cal W}_T + {\cal W}_L)\, .
\label{eq: unpol xsec}
\end{equation}
 It is then useful to introduce the ratio of polarized and unpolarized cross sections
\begin{align}
\frac{\d N}{\d \Omega} \equiv \frac{\d\sigma}{\d \xB\,  \d y\,\d z\, \d^2 \bm P_T\,\d \Omega } \left ( \frac{\d\sigma}{\d \xB\,  \d y\,\d z\, \d^2 \bm P_T}  \right )^{-1}\,,
\end{align}
which can be expressed as follows
\begin{align}
\frac{\d N}{\d \Omega} = \frac{3}{4 \pi}\, \frac{1}{\lambda + 3}\, \left [1 + \lambda \cos^2\theta + \mu  \sin 2\theta \cos\varphi  +  \frac{1}{2} \, \nu\,   \sin^2\theta \cos 2\varphi \right ]\,,
\label{eq: pol parameterization 2}
\end{align}
where we have defined the polarization parameters
\begin{equation}
\lambda = {{\cal W}_{1 1} - {\cal W}_{0 0} \over {\cal W}_{1 1} + {\cal W}_{0 0}}, \quad \mu = {\sqrt 2\, \Re\left[ {\cal W}_{1 0} \right] \over {\cal W}_{1 1} + {\cal W}_{0 0}}, \quad \nu = {{\cal W}_{1,\,-1} \over  {\cal W}_{1 1} + {\cal W}_{0 0}}\, ,
\label{eq: lmn parameters}
\end{equation}
or alternatively adopting Eq.~\eqref{eq:W-P-L},
\begin{equation}
\lambda = {{\cal W}_T - {\cal W}_L \over {\cal W}_T + {\cal W}_L}, \quad \mu = {{\cal W}_\Delta \over {\cal W}_T + {\cal W}_L}, \quad \nu = {2\, {\cal W}_{\Delta\Delta} \over {\cal W}_T + {\cal W}_L} \, .
\end{equation}
The parameterizations shown in Eqs.~\eqref{eq: polarization parameterization 1} and \eqref{eq: pol parameterization 2} are standard for the study of the angular distribution of a spin-one particle decay into a lepton pair and, indeed, they are commonly adopted in Drell-Yan processes~\cite{Lam:1978pu} and in $J/\psi$ photoproduction~\cite{Beneke:1998re}.

Among the polarization coefficients, $\lambda$, $\mu$ and $\nu$, the most investigated experimentally is $\lambda$.
Moreover, from the phenomenological point of view it has a very intuitive interpretation, with $\lambda=+1(-1)$ describing a \textit{transverse}(\textit{longitudinal}) polarization state for the $J/\psi$ (\textit{i.e.}~a $J/\psi$ helicity equal to $\pm 1$ or 0), while $\lambda = 0$  for an \textit{unpolarized} one.

The main goal of our study is to present estimates for these polarization quantities, within both the CSM and the NRQCD frameworks, focusing on the kinematic region accessible at the future EIC. As we will show in the following, such a detailed phenomenological study could help in disentangling among the production mechanisms.

\begin{table}[t]
	\centering
	\begin{tabular}{|c|||c||c|c|c|}
		\hline
		\rule[-1.5ex]{0pt}{4.5ex} {\large LDME Set } & {\large $ \braket{{\cal O}_1[\,^3 S_1]}$} $\big[{\rm GeV}^3 \big]$ & {\large $ \braket{{\cal O}_8[\,^1 S_0]}$} $\big[{\rm GeV}^3 \big]$ & {\large $ \braket{{\cal O}_8[\,^3 S_1]}$} $\big[{\rm GeV}^3 \big]$ & {\large $ \braket{{\cal O}_8[\,^3 P_0]}$} $\big[{\rm GeV}^5 \big]$  \\
		\hline
		\rule[-1ex]{0pt}{3.5ex} C12 & $1.16$ & $0.089$ & $0.003$ &$ 0.0126$ \\
		\hline
		\rule[-1ex]{0pt}{3.5ex} G13 & $1.16$ & $0.097$ & $-0.0046$ & $-0.0214$ \\
		\hline
		\rule[-1ex]{0pt}{3.5ex} BK11 & $1.32$ & $0.0304$ & $0.00168$ & $-0.00908$ \\
		\hline
	\end{tabular}
	\caption{LDME set (central) values for the $J/\psi$ state: C12~\cite{Chao:2012iv},
	G13~\cite{Gong:2012ug} and BK11~\cite{Butenschoen:2011yh}. For the other $^3 P_J$ states we use the standard spin-symmetry relation
	$ \braket{{\cal O}_8[\,^3 P_J]} = (2J+1)\, \braket{{\cal O}_8[\,^3 P_0]}$. }
	\label{tab: LDME sets for Jpsi}
\end{table}

\section{Angular distributions}
\label{sec: ang_distr}

In this section we analyze the polarization parameters defined in Eq.~(\ref{eq: lmn parameters}) showing both their $z$ and $P_T$ distributions. The explicit analytic expressions of the underlying partonic structure functions, calculated at the perturbative order $\alpha_s^2$, are presented in Ref.~\cite{DAlesio:2021yws} for the so-called \textit{Gottfried-Jackson} frame, together with all prescriptions needed to transform them in the other relevant frames. For the predictions based on the NRQCD approach, in addition to the CS contribution, given by a pure gluon fusion channel, we consider the CO channels up to the order $v^4$, which involve both gluon and quark final states. The CTEQ6L1 set~\cite{Pumplin:2002vw} is used for the unpolarized parton distribution functions. Moreover, in order to assess the stability of our results against higher order corrections, we produce uncertainty bands by varying the factorization scale $\mu_{F}$ in the range $ \mu_0 / 2 < \mu_{F} < 2 \mu_0 $, around the central value $\mu_0 = \sqrt{Q^2 + M_\psi^2}$.

Concerning the CO LDME values, three different sets are adopted, see Table~\ref{tab: LDME sets for Jpsi}. Here we only recall their main features: the C12 set~\cite{Chao:2012iv} has been extracted simultaneously from both polarized and unpolarized $J/\psi$ production data in $pp$ collision at $P_T>7$~GeV, measured by the CDF (Run II) Collaboration; the G13 set~\cite{Gong:2012ug} is obtained including only $P_T>7$~GeV unpolarized data from the CDF and LHCb Collaborations and then used to predict the $J/\psi$ polarization in $pp$~collisions; it is in agreement with the C12~set if feed-down contribution is negligible; the BK11 set~\cite{Butenschoen:2011yh} is based on a fit without any polarization data, but starting from a lower $P_T$ value, around $3$~GeV, and including both photoproduction and hadroproduction data.

The high cm energy kinematical set-ups expected at the EIC are an ideal environment to study $J/\psi$ polarization in electroproduction. Moreover, they will allow to better explore high photon virtualities ($Q$), avoiding the competing contributions from photoproduction.
Furthermore, since we are interested in the region where collinear factorization holds, our results will be shown only for $P_T$ values above $P_{T\rm {min}} = 1$~GeV. Notice that around this value we actually enter the region where the transverse momentum dependent (TMD) factorization could be applied and therefore our estimates are pushed down to the overlapping region of validity of the two factorization schemes.

\subsection{The $\lambda$ parameter}

\begin{figure}[t]
	\centering
	\includegraphics[width= 1\linewidth, keepaspectratio]{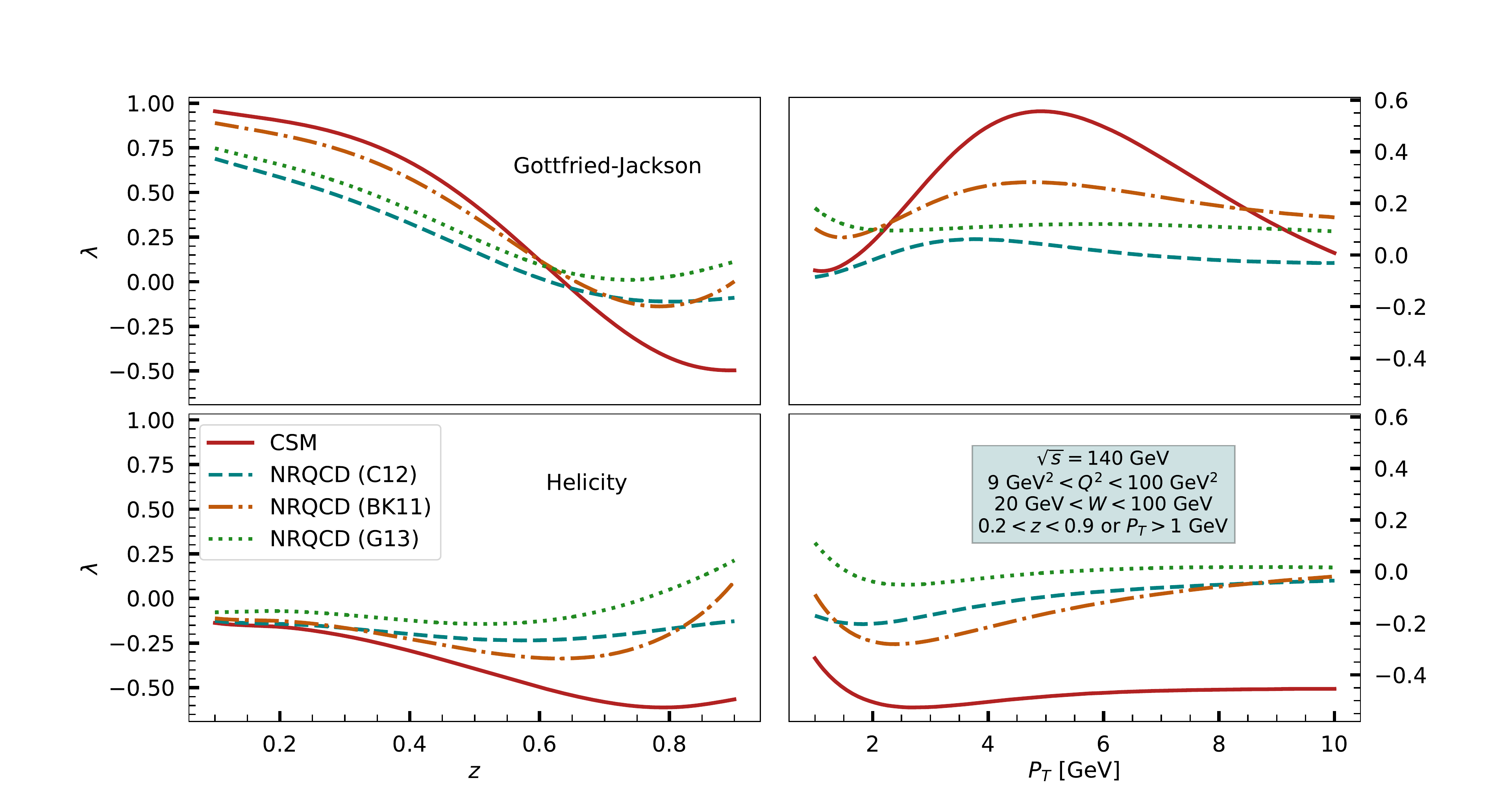}
	\caption{Estimates for $\lambda$ at $\sqrt s=140$~GeV as a function of $z$ (left panels) and $P_T$ (right panels) for different models and LDME sets and two reference frames: \textit{Gottfried-Jackson} (upper panels) and \textit{Helicity} (lower panels) frames. Integration ranges are given in the light-blue legend box.}
	\label{fig: lambda at 140GeV}
\end{figure}

In Fig.~\ref{fig: lambda at 140GeV} we present our predictions for $\lambda$ at $\sqrt s=140$~GeV, as a function of both the $J/\psi$ energy fraction $z$ (left panels) and its transverse momentum $P_T$ (right panels). Two quarkonium rest frames are explicitly considered: the \textit{Gottfried-Jackson} (upper panels) and the \textit{Helicity} (lower panels) ones. In this and in the following figures, the kinematical ranges explored are indicated in the legend boxes. For completeness we report here the corresponding regions explored in $x_B$ and $y$ at $\sqrt s=140$~GeV, $10^{-3} \lesssim x_B \lesssim 0.2$ and $y \lesssim 0.5$ respectively, even if the effectively probed maximum value in $x_B$ is around 0.07.

Concerning other typical frames, like the \textit{Target} and \textit{Collins-Soper} ones, we only notice that the first one give estimates very close to those in the \textit{Helicity} frame, while predictions obtained in the second one, at least for the kinematics considered, are in general much smaller than those in the \textit{Gottfried-Jackson} frame or even close to zero.

Notice that for such observable, defined as a ratio of cross sections, the dependence on the scale $\mu_{F}$ in the range $[\mu_0/2, 2 \mu_0]$ is barely appreciable and therefore is not shown.

\begin{figure}[ht]
	\centering
	\includegraphics[width= 1\linewidth, keepaspectratio]{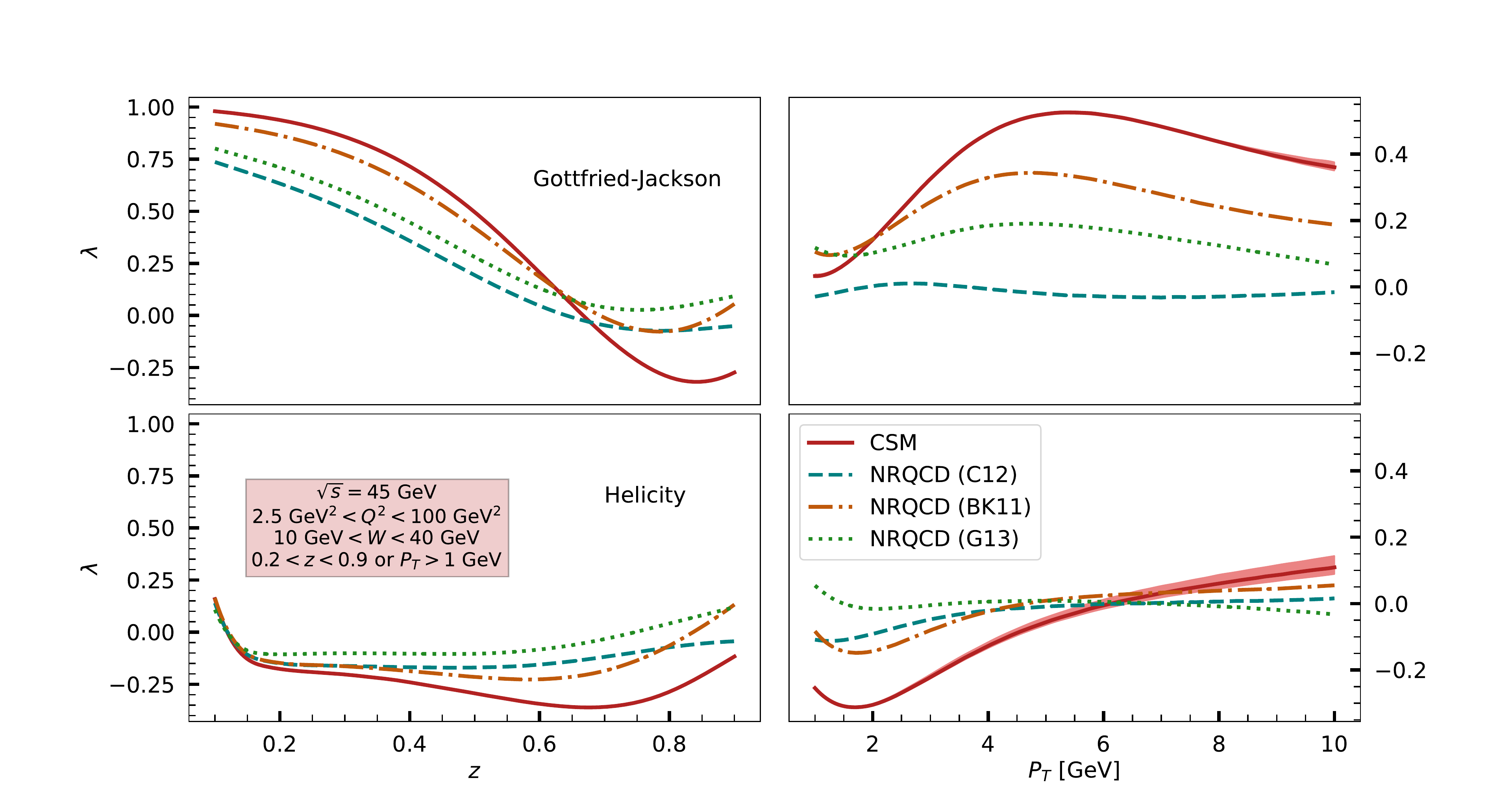}
	\caption{
	Estimates for $\lambda$ at cm~energy $\sqrt s = 45$~GeV. The integration region, different with respect to the higher-energy case, is given in the red legend box, while curves and panels have the same meaning as in Fig.~\ref{fig: lambda at 140GeV}. The scale error bands are sizable and explicitly shown only for the CSM prediction as a function of $P_T$.}
\label{fig: lambda at 45GeV}
\end{figure}

The study of the $\lambda$ parameter as a function of $z$ presents very interesting features from the phenomenological point of view. The reasons are manifold: first of all its expected relative large size as compared to the $\mu$ and $\nu$ parameters.
Moreover, it is experimentally under more active investigation. On the other hand, theoretical estimates for $\lambda$ as a function of $z$ (for small and moderate values) do not vary significantly adopting different frameworks (Fig.~\ref{fig: lambda at 140GeV}, left panels), which implies that, in order to get information on the quarkonium formation mechanism, one would need highly precise measurements. The same problem was found in different analyses performed by the HERA Collaborations, Refs.~\cite{H1:2002xeb,Yuan:2000cn}.

The situation changes considerably at $z>0.6$, which represents a very interesting region from the phenomenological point of view. As is well known, NRQCD estimates for the unpolarized cross section manifest a divergent behavior as $z \to 1$, due to the corresponding $\hat t \to 0$ singularities.
This can potentially spoil the validity of NRQCD~factorization. As shown in Ref.~\cite{Beneke:1999gq}, in order to extend the region of applicability of NRQCD up to $1 - z \sim v^2$, one can introduce a new set of functions, the so-called shape functions~\cite{Beneke:1997qw}, that allow to improve noticeably the convergence for photoproduction. We expect such quantities to be relevant also for the SIDIS process, together with their TMD extensions, which have been adopted in the study of $pp$ collisions in Refs.~\cite{Echevarria:2019ynx,Fleming:2019pzj} and whose perturbative tails have been derived in Ref.~\cite{Boer:2020bbd} for unpolarized and in Ref.~\cite{DAlesio:2021yws} for polarized $J/\psi$ SIDIS. On the other hand, the impact of the shape functions on $\lambda$ is expected to be strongly reduced since $\lambda$ is a ratio of cross sections. This can be tested with future available data.

A much more powerful tool to assess the relevance of the CO contributions is the study of the $P_T$ distribution (Fig.~\ref{fig: lambda at 140GeV}, right panels).
In the \textit{Gottfried-Jackson} frame (upper panel) we see a clear separation as well as a different behavior between the CSM and NRQCD curves, in particular in the region $4<P_T<7$~GeV; similarly in the \textit{Helicity} frame there is a wide separation between the CSM and the NRQCD curves, while different LDME sets give predictions much closer to each other and closer to $\lambda = 0$.
It is worth noticing that, even if the unpolarized cross section decreases as $P_T$ increases, a good separation can be found already around $P_T\simeq 5$~GeV, which is also far away from the TMD region.

Before concluding the analysis of $\lambda$ at large cm~energies, a comment on the contributions from different partonic channels and/or different NRQCD waves can be useful.
Concerning the $z$ distribution, we find that the main contribution to the numerator of $\lambda$ comes from the (gluon) CS~wave, while the differences among NRQCD predictions, especially around $z \to 0.9$, are due to the gluon \textit{P}-wave, modulated by the corresponding LDME parameter.
For the $P_T$ distribution we find, similarly, that the CS~term is on the whole the most relevant contribution, followed again by the gluon \textit{P}-wave one.
In particular at $P_T \to 1$~GeV the size of the gluon \textit{P}-wave contribution becomes comparable to (or even larger than) the CS one; moreover,
since the low-$P_T$ region dominates the integration over $P_T$, one can also understand why the gluon \textit{P}-wave is so relevant in our estimates vs.~$z$, with the most visible effects for $z \to 0.9$.

At medium $P_T$ values the quark \textit{P}-wave starts becoming important and at even higher $P_T$ values it is similar in size to the gluon one; this means that in this region, the full \textit{P}-wave contribution (gluon+quark) dominates over the CS one.

Another interesting possibility given by the future EIC facility is the corresponding analysis at smaller energies: in the following we will adopt $\sqrt s = 45$~GeV.
In this case, different integration ranges have been considered for $W$ and $Q^2$, as reported in the legend box of Fig.~\ref{fig: lambda at 45GeV}. These, in turn, correspond to $10^{-3}\lesssim x_B \lesssim  0.5 $ (with an effective upper limit around $x_B\simeq 0.2$) and $y\lesssim 0.8$, a more valence-like region w.r.t.~the previous case. Moreover, since at lower energies it is more difficult to reach high photon virtualities, we get contributions mostly from moderately low $Q^2$. Consistently we adopt a lower limit, $Q_{\rm min}\simeq 1.6$~GeV, in the integration.
Notice that in this kinematic region, at least for the high $P_T$ dependence of $\lambda$ within the CSM, the scale error bands are once again sizeable enough.

From Fig.~\ref{fig: lambda at 45GeV} (left panels) we can see that the $z$ distribution does not depend significantly on the energy for $z \leq 0.6$, while at higher $z$ values the estimates are closer to zero, at variance with those at higher cm~energy. As said, a polarization study pushed up to this regime can suffer from factorization breaking effects in NRQCD even if data in this region could be relevant from the phenomenological point of view.
We also observe a rapid variation of all curves in the \textit{Helicity} frame at $z\sim 0.1$. This is due to geometrical factors which are energy dependent (see also Eq.~(A16) of  Ref.~\cite{DAlesio:2021yws}).
The same variation is also present at higher cm~energy, but for $z < 0.1$ (outside the range shown in the lower-left panel of Fig.~\ref{fig: lambda at 140GeV}).

Concerning the $P_T$ dependence, Fig.~\ref{fig: lambda at 45GeV} (right panels), we notice that the CSM results are very different with respect to the corresponding ones in Fig.~\ref{fig: lambda at 140GeV}, while the same is not true for the NRQCD cases.
This is related to the different virtualities explored, on which the CSM estimates depend heavily.
This difference can be considered as an extra tool in the quest of discerning among different frameworks.

Finally, we briefly comment on how the parton and/or wave contributions vary with the energy. While the $z$ distribution manifests almost no energy dependence, the $P_T$ spectrum presents interesting features in the two frames considered. For the \textit{Gottfried-Jackson} one the relative contribution from the quark \textit{P}-wave is widely increased at this lower energy, making it the leading term in the numerator at medium/high $P_T$. Regarding the \textit{Helicity} frame the situation is, potentially, even more interesting, since the CSM and \textit{P}-wave (both gluon and quark) contributions are highly suppressed at this energy, especially at large $P_T$. The main role is then played by the $^3\! S_1^{(8)}$~quark wave, which is responsible for the difference among the predictions based on the LDME sets considered. Even if in this region it is quite hard to expect precise enough data to discriminate between models, it is nevertheless worth stressing that it could be very useful in constraining the nonperturbative physics.

\subsection{The $\mu$ parameter}

\begin{figure}[t]
\centering
\includegraphics[width= 1\linewidth, keepaspectratio]{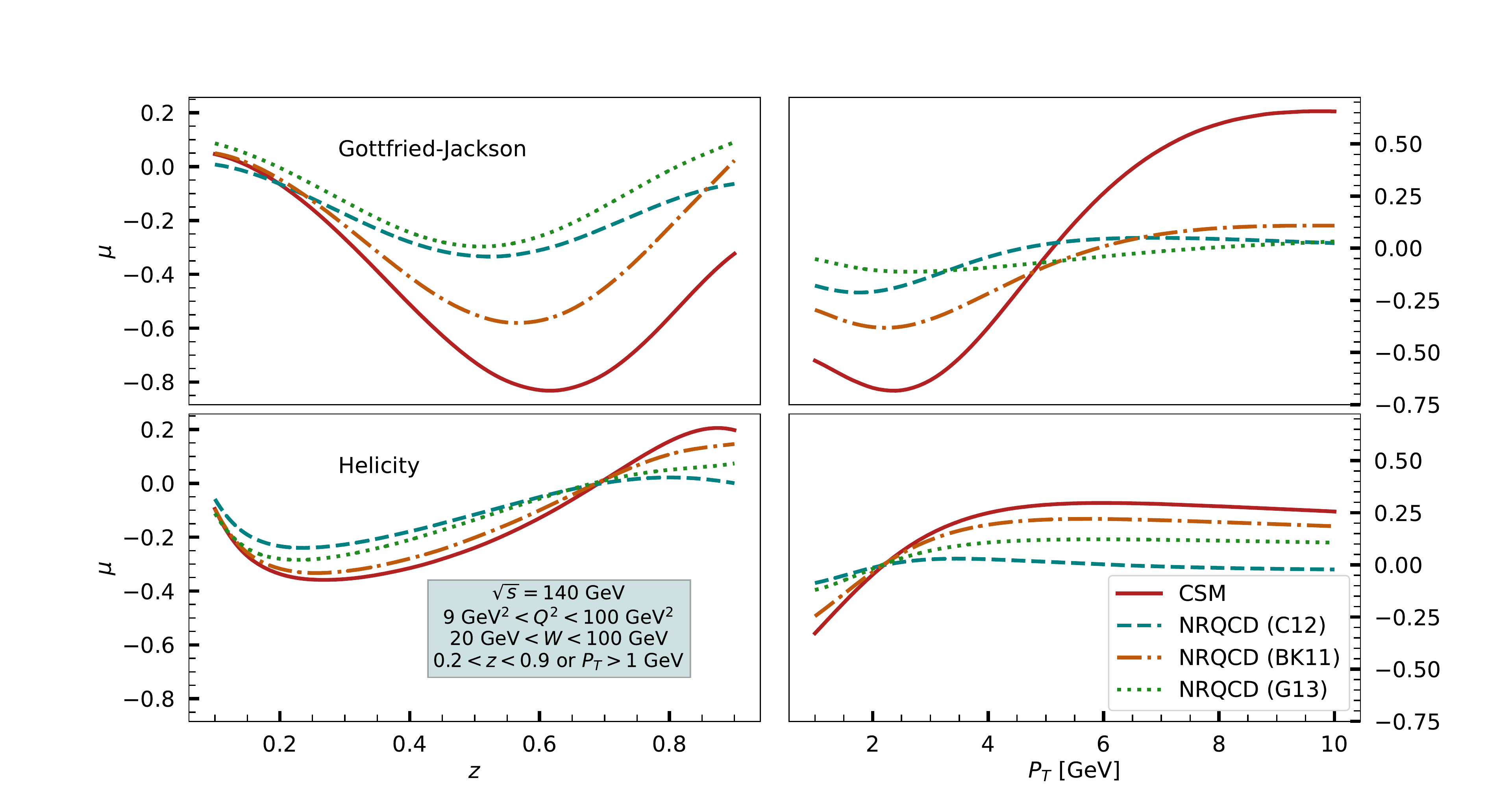}
\caption{Estimates for the parameter $\mu$ at $\sqrt s = 140$~GeV. Paneling order is the same as in Fig.~\ref{fig: lambda at 140GeV}. Integration ranges are given in the blue legend box.}
\label{fig: mu at 140GeV}
\end{figure}

\begin{figure}[ht]
\centering
\includegraphics[width= 1\linewidth, keepaspectratio]{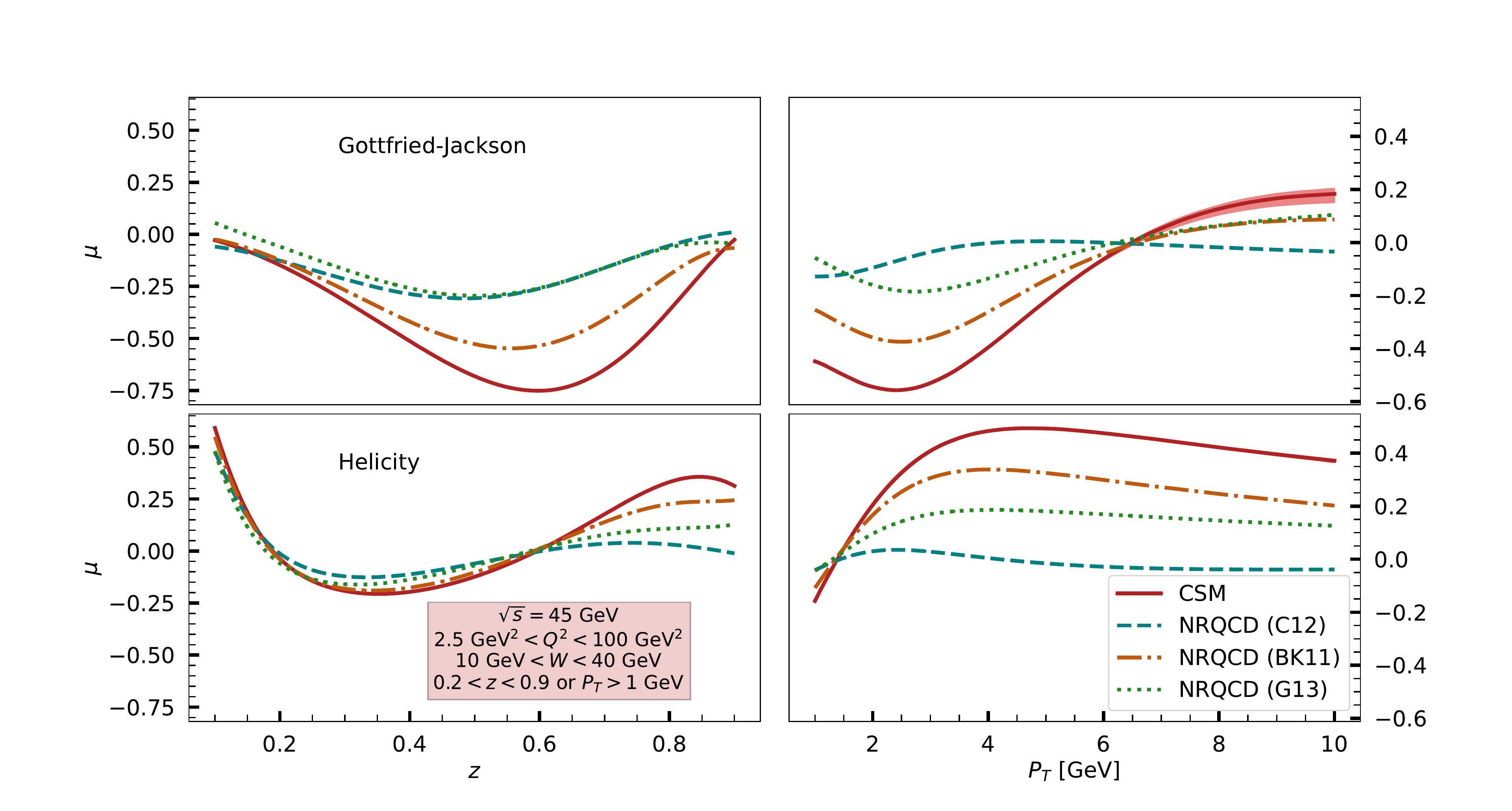}
\caption{Estimates for the parameter $\mu$ at $\sqrt s = 45$~GeV. Paneling order is the same as in Fig.~\ref{fig: lambda at 140GeV}. Integration ranges are given in the red legend box.}
\label{fig: mu at 45GeV}
\end{figure}

Estimates for the $\mu$ parameter are again provided both in the \textit{Gottfried-Jackson} and in the \textit{Helicity} frames, as a function of $z$ and $P_T$ at $\sqrt s = 140$~GeV, Fig.~\ref{fig: mu at 140GeV}, and $\sqrt s = 45$~GeV, Fig.~\ref{fig: mu at 45GeV}.

From these figures we see that the \textit{Gottfried-Jackson} frame is the best choice to discern among the CSM and NRQCD approach. A similar conclusion holds for the parameter $\nu$ as well, see the discussion in Sec.~\ref{sec:nu}.
Indeed, in Fig.~\ref{fig: mu at 140GeV} the separation between the CSM estimates and the corresponding NRQCD ones are remarkably sizeable for $z \gtrsim 0.5$ and $P_T \gtrsim 5$~GeV.
On the contrary, estimates in the \textit{Helicity} frame both with respect to $z$ and $P_T$ are so close to each other that one cannot draw any conclusion.

The wave/parton decomposition of the ${\cal W}_\Delta$ helicity function, that is directly related to the $\mu$ numerator,  allows us to get some further insights. The main CO contribution comes from the $P$-wave term. In particular, differences in NRQCD predictions as a function of $z$ (left panels of Fig.~\ref{fig: mu at 140GeV}) are driven by the gluon $P$-wave LDMEs. Moreover, the gluon $P$-wave dominates the numerator behavior with respect to $P_T$ too (right panels of Fig.~\ref{fig: mu at 140GeV}). In addition, we find that the NRQCD predictions in the \textit{Gottfried-Jackson} frame receive a significant contribution from the gluon $P$-wave also at low-$P_T$, namely $P_T \lesssim 3$~GeV.
At variance with the behavior in $z$, here the quark $P$-wave channel is relevant at high $P_T$, especially when considering the \textit{Helicity} frame.

Moving to the lower cm energy, we see that the CSM $\mu$ estimates in the \textit{Gottfried-Jackson} frame, Fig.~\ref{fig: mu at 45GeV} (upper panels), vary significantly for $z \gtrsim 0.5$ and $P_T \gtrsim 5$~GeV, as compared with what happens at $\sqrt s = 140$~GeV.
We remark that this variation can also appear via a proper $Q$-binning in the higher cm energy case ($\sqrt s = 140$~GeV).
In contrast, estimates within the \textit{Helicity} frame at lower energies (lower panels of Fig.~\ref{fig: mu at 45GeV}) do not present the same energy/$Q$-binning dependence. The only remarkable exception resides in the $P_T$ distribution, where CSM predictions increase up to $\sim 40\%$, to be compared with the $\sqrt s = 140$~GeV case where the CSM result is at most $\sim 25\%$. Despite this, $\mu$ estimates in the \textit{Helicity} frame do not differ enough to discern among different models.

Looking at the wave/parton decomposition, we confirm that also for the $\mu$ numerator the role of quarks is enhanced at lower energies. This is particularly true for the $P_T$ dependence. Here we find that NRQCD predictions at the higher $P_T$ values, namely $P_T \gtrsim 6$~GeV, are mostly driven by the quark $P$-wave; moreover, in the same $P_T$ region we observe that the $^3 S_1^{[8]}$ quark wave is non-negligible.

\subsection{The $\nu$ parameter}
\label{sec:nu}
We now discuss the parameter $\nu$, which is particularly important in the TMD framework, since it is directly related to the TMD distribution of linearly polarized gluons inside an unpolarized proton, $h_1^{\perp g}$.
This could play a role in the region of moderately low $P_T$, where the two factorization schemes overlap.

\begin{figure}[t]
\centering
\includegraphics[width= 1\linewidth, keepaspectratio]{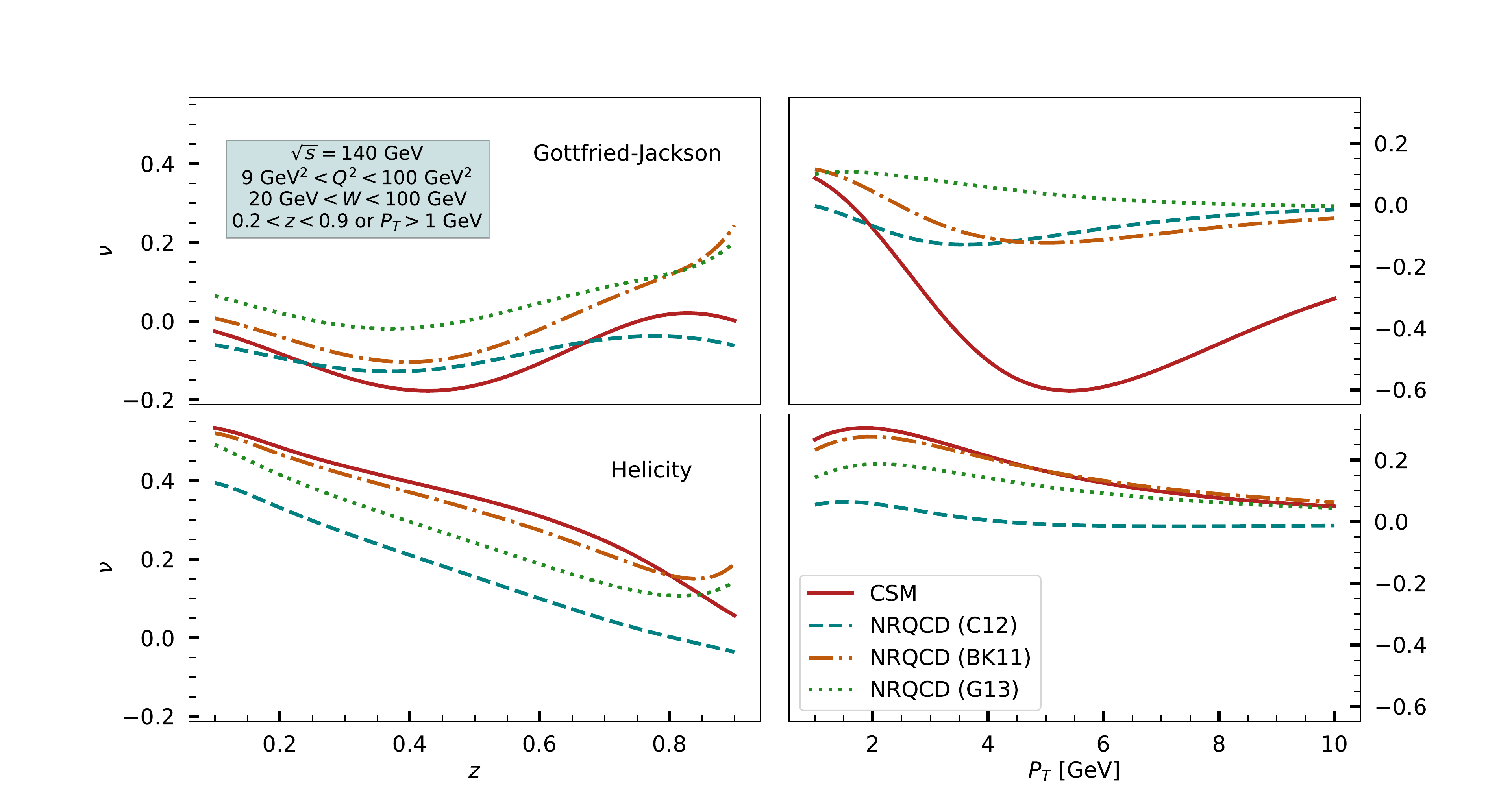}
\caption{Estimates for the parameter $\nu$ at $\sqrt s = 140$~GeV. Paneling order is the same as in Fig.~\ref{fig: lambda at 140GeV}. Integration ranges are given in the blue legend box.}
\label{fig: nu at 140GeV}
\end{figure}

\begin{figure}[t]
\centering
\includegraphics[width= 1\linewidth, keepaspectratio]{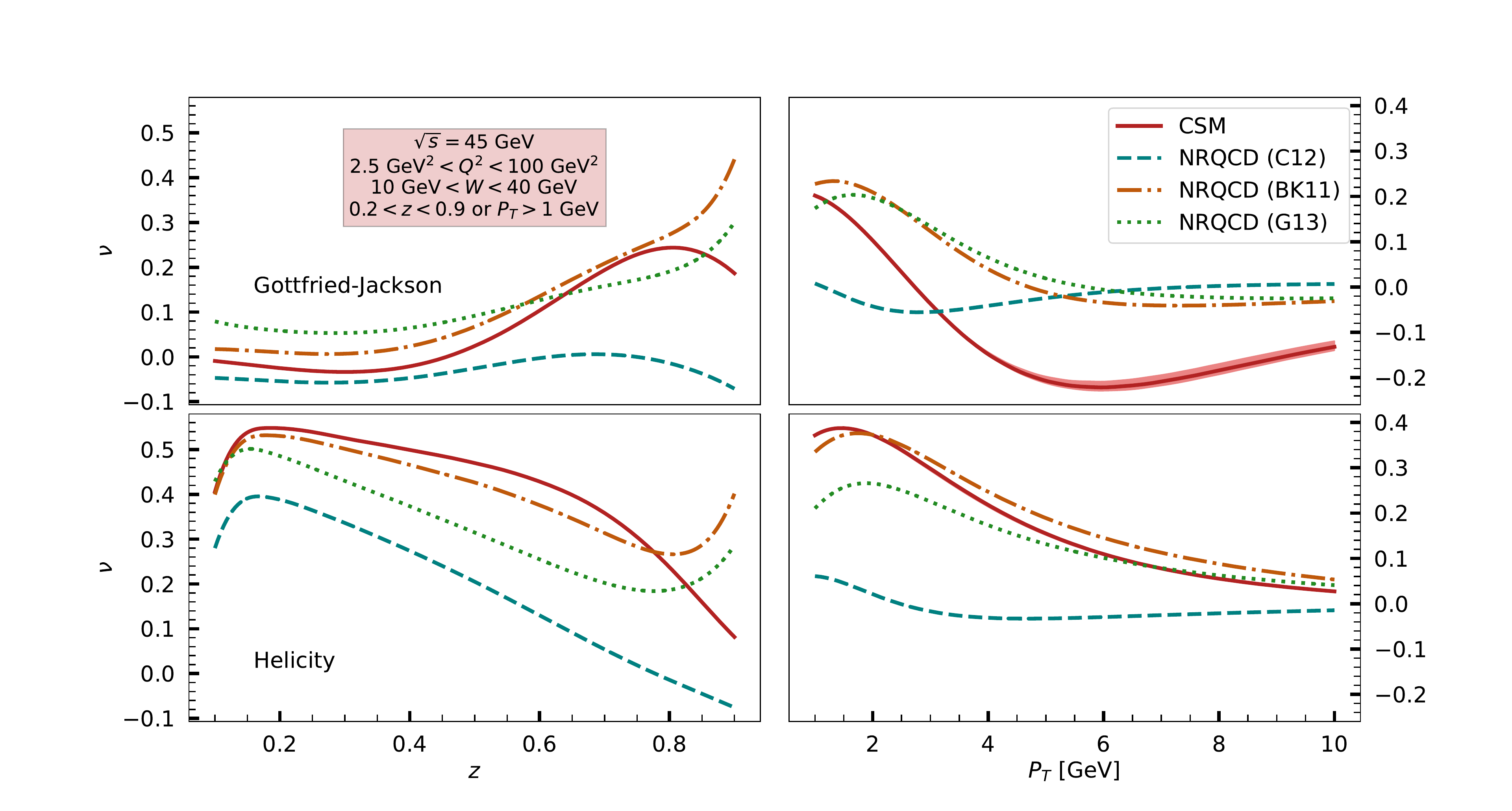}
\caption{Estimates for the parameter $\nu$ at $\sqrt s = 45$~GeV. Paneling order is the same as in Fig. \ref{fig: lambda at 140GeV}. Integration ranges are given in the red legend box.}
\label{fig: nu at 45GeV}
\end{figure}

Again, we focus initially on the higher cm~energy ($\sqrt s = 140$~GeV), Fig.~\ref{fig: nu at 140GeV}, and then we describe the main differences with respect to the smaller cm~energy ($\sqrt s = 45$~GeV),  Fig.~\ref{fig: nu at 45GeV}.

Starting from the $z$-dependent distribution in Fig.~\ref{fig: nu at 140GeV} (left panels), we see once again that even if the estimated $\nu$ values are potentially sizeable, at least in the \textit{Helicity} frame, the separation among the different approaches is in general very poor.
Nevertheless, it is worth remarking that at high~$z$ we find more sensitivity to the LDME sets in the NRQCD framework.
The situation is slightly different for the $P_T$ case (right panels): if the \textit{Helicity} frame does not show a promising scenario, in the \textit{Gottfried-Jackson} case the differences in the medium/high-$P_T$ region between the two approaches are sizeable.

As said, results at high~$z$ and/or small~$P_T$ are in general promising for future analyses regarding the $h_1^{\perp g}$ gluon distribution in the TMD region.
Nevertheless, it is important to remark that for the $\nu$ parameter the shape functions and their TMD extensions enter, potentially, in a different way in the numerator and the denominator, and their role could be important. This requires further investigation, together with a full higher-order description in $\alpha_s$, which is not available at present.

It is once again interesting to look into the parton and wave decomposition. The $z$-dependent ${\cal W}_{\Delta\Delta}$ is dominated, for almost all $z$ values, by the CS~wave; only for $z \to 0.9$ the CS~contribution becomes negligible, and the results are driven by the CO \textit{P}-wave, in particular by the gluon term.
Moving to the $P_T$ dependence, we find again some similarities with the $\lambda$ case: the CS term is the relevant contribution to the numerator over the whole $P_T$ spectrum, together with the gluon \textit{P}-wave. At variance with the $\lambda$ parameter case, the quark contribution to the \textit{P}-wave term starts becoming important already at small-$P_T$ values.

Moving to the lower cm~energy, from Fig.~\ref{fig: nu at 45GeV} we see that the $z$ distribution is sensitive to the energy change in the whole spectrum, at variance with the $\lambda$ case. The differences, particularly noticeable in the \textit{Gottfried-Jackson}~frame, are mostly in size and not in the general behavior, implying that even in this case it would be difficult to extract any information. Again, we remark that the rapid variation of $\nu$ estimates at low-$z$ values is due to a geometrical factor (Eq.~(A16) of Ref.~\cite{DAlesio:2021yws}).
The $P_T$-dependent distributions, instead, have a quite different behavior for the two frames displayed.
The \textit{Gottfried-Jackson} estimates vary significantly in size, especially if one considers the CSM; moreover all the LDME sets give similar predictions, compatible with zero, for $P_T > 5$~GeV, while predictions, in both approaches, are sizeable (up to $\sim 20\%$) at low-$P_T$ values. This could be very promising for further extensions to the TMD~region.
The curves in the \textit{Helicity}~frame, instead, do not show the same dependence on the energy. In general, we conclude that the study of the $\nu$ parameter, at least in this frame, is not very effective. Nevertheless it becomes more interesting when its information is combined with other parameters, as done in the study of the invariant quantities in the next section, Sec.~\ref{sec: invariants}.

Concerning the wave decomposition, we find that both quark and gluon \textit{P}-wave contributions to the $P_T$ and $z$ distributions are enhanced at lower energies, even if for the latter this is true only at large $z$ values.
Notice that the different (larger) size of the $\nu$ parameter at $z \to 0.9$ could also affect the TMD region, increasing the possibility of extracting information on the linearly polarized gluon distribution. The main source of this enhancement at $\sqrt s = 45$~GeV is related once again to the lower photon virtualities explored. In this sense, very similar predictions might be expected at higher cm~energy via a binned analysis with $1.6~{\rm GeV}<Q<M_\psi$.

\section{Rotational invariants}
\label{sec: invariants}
The polarization parameters $\lambda$, $\mu$ and $\nu$, as widely discussed in the previous sections, are frame dependent by definition, since they are expressed with respect to the solid angle $\Omega$ spanned by the $l^+$ particle in the $J/\psi$ decay and in its rest frame. As already pointed out, the frame choice is not unique and the results appear different from frame to frame. On the other hand, the relations among the most used reference frames are computable, since they differ only in the $Z$-axis direction.

A complementary and powerful tool to study $J/\psi$ polarization, both from the experimental and the phenomenological points of view, is the use of rotational invariant parameters, that are rest-frame independent by construction. These can be defined taking into account what follows.

For all the most common choices, the $Z$- and $X$-axes, lying in the $J/\psi$ production plane, are defined in terms of physical momenta in the quarkonium rest frame (see Appendix A of Ref.~\cite{DAlesio:2021yws}), with the $Y$-axis always perpendicular with respect to this plane and always pointing in the same direction. This implies that two frames ($F,F'$) can be connected by a simple rotation of an angle $\psi$ around the $Y$-axis, and the corresponding polarization parameters can be directly related as\footnote{Here $\mu_F$ stands for the $\mu$ parameter in a specific frame $F$, not to be confused with the factorization scale $\mu_F$ defined in the previous sections.}
\begin{equation}
\begin{pmatrix}
	\lambda \\ \mu \\ \nu
\end{pmatrix}_{F'}
=
{1 \over 1 +\rho}
\begin{pmatrix}
	1 - {3 \over 2} \sin^2 \psi & {3 \over 2} \sin 2\psi  & {3 \over 4} \sin^2 \psi \\
	-{1 \over 2}\sin 2\psi & \cos 2\psi & {1 \over 4} \sin 2\psi \\
	\sin^2 \psi & -\sin 2\psi & 1 - {1 \over 2} \sin^2 \psi
\end{pmatrix}
\begin{pmatrix}
	\lambda \\ \mu \\ \nu
\end{pmatrix}_F
\, ,
\label{eq: lmn rotation matrix}
\end{equation}
with
\begin{equation}
\rho =  {\sin^2 \psi \over 2} \left( \lambda_F - {\nu_F \over 2} \right) - \sin 2\psi\,{\mu_{F} \over 2} ,
\end{equation}
as given in Eqs.~(A.18) and (A.19) of Ref.~\cite{DAlesio:2021yws}, where we have changed the rotation angle from $\theta$ to $\psi$ to avoid any confusion with the polar angle of the final lepton $l^+$. Notice that the quantity $\rho$ depends on the kinematics, since the rotation angle itself depends on the partonic Mandelstam variables 
(see Eqs.~(A.14)-(A.16) of Ref.~\cite{DAlesio:2021yws} for details).

From Eq.~\eqref{eq: lmn rotation matrix}, one can construct several quantities which do not change upon rotation around the $Y$~direction. The following relations are extremely useful in this respect:
\begin{equation}
    3 + \l_{F'} = {1 \over 1 + \rho} \left( 3 + \l_F \right), \qquad 1 - {\nu_{F'} \over 2} = {1 \over 1 + \rho} \left( 1 - {\nu_F \over 2} \right) \,.
\end{equation}
A group of rotational invariants, as initially proposed in Ref.~\cite{Faccioli:2010ji}, can be defined in terms of two polarization parameters, namely $\lambda$ and $\nu$,
\begin{equation}
	{\cal F}_{(c_i)} = {c_0 (3 + \lambda) + c_1 (1 - \nu/2) \over c_2 (3 + \lambda) + c_3 (1 - \nu/2)}  \, ,
	\label{eq: invariant F}
\end{equation}
where $c_i$ are suitable free constants.

Among all possible combinations, two of them play an important role and have received special attention~\cite{Faccioli:2010ej, Faccioli:2010kd, Faccioli:2011zzb,Faccioli:2011zzc,Peng:2018tty}
\begin{equation}
  {\cal F} \equiv {\cal F}_{(1, -2, 1, 0)} = {1 + \lambda + \nu \over 3 + \lambda}
	\label{eq: invariant F1}
\end{equation}
and
\begin{equation}
    \tilde \lambda \equiv {\cal F}_{(1, -3, 0, 1)}  = {2\, \lambda + 3\, \nu \over 2 - \nu} \,.
	\label{eq: invariant F2}
\end{equation}
These invariants have been widely studied in $pp$ and heavy-ion processes~\cite{Ma:2018qvc,ALICE:2018crw}.

It is worth noticing that both invariants can be similarly defined for Drell-Yan processes, where they acquire a constant value if the Lam-Tung relation ($1 - \lambda = 2 \nu$) holds~\cite{Lam:1978pu}: ${\cal F}_{\rm DY} = 1/2$ and $\tilde \lambda_{\rm DY} = +1$, as pointed out in Refs.~\cite{Faccioli:2010kd,Peng:2018tty}. Another interesting feature is that $\tilde \lambda = +1(-1)$ is related to a natural transverse (longitudinal) polarization~\cite{Faccioli:2010ji}.
It is important to stress that the constant behavior is purely dynamical, and in particular for the Drell-Yan case is a consequence of rotational invariance and helicity conservation~\cite{Faccioli:2011pn}. Since $J/\psi$ couples differently in SIDIS processes, the Lam-Tung relation is expected to be broken in this case.

\begin{figure}[t]
\centering
\includegraphics[width= 1\linewidth, keepaspectratio]{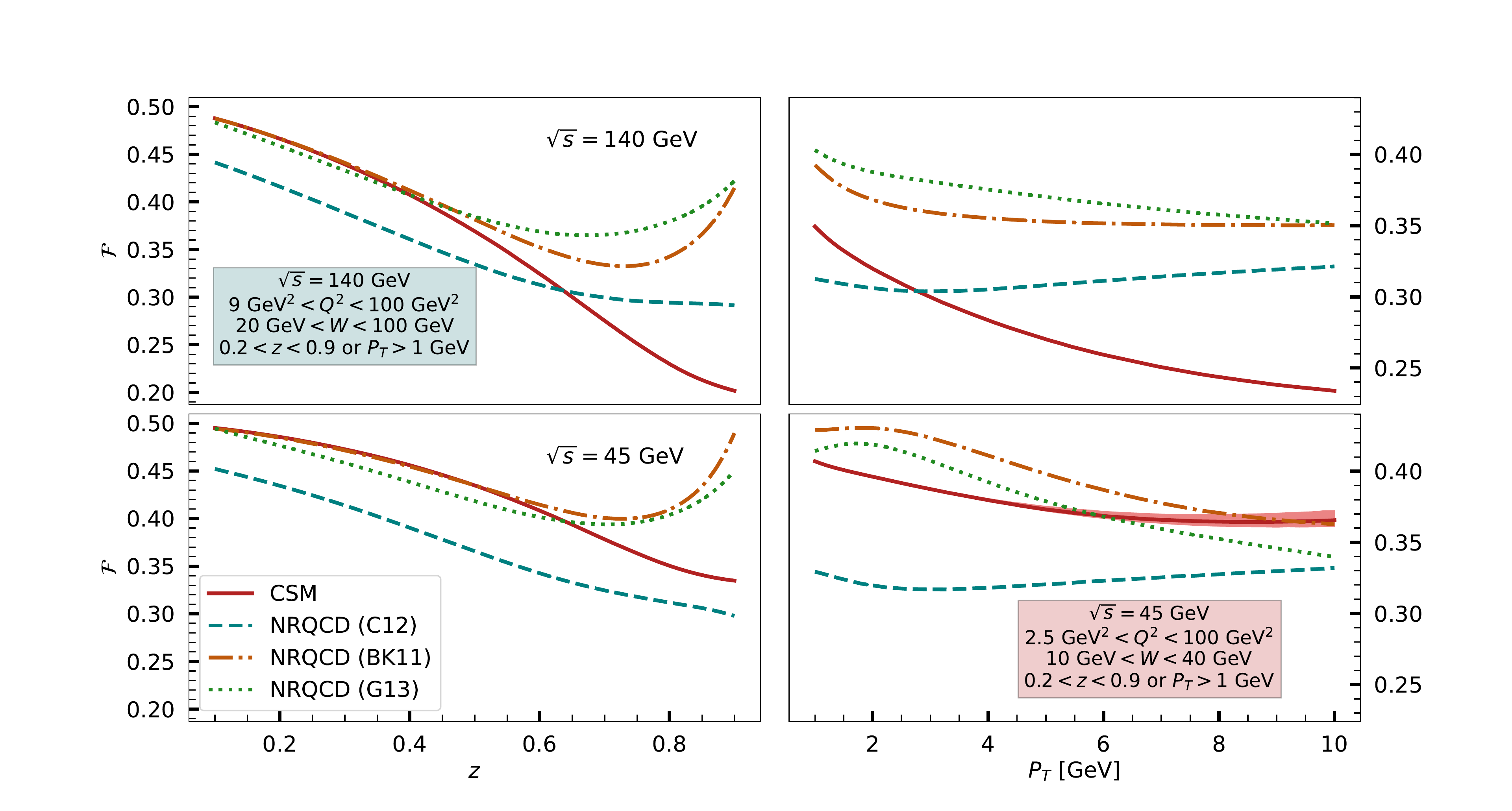}
\caption{Estimates for the invariant ${\cal F}$, Eq.~({\ref{eq: invariant F1}}), as a function of $z$ (left panels) and $P_T$ (right panels) at two cm energies, $\sqrt s = 140$~GeV (upper panels) and $\sqrt s = 45$~GeV (lower panels), for different approaches and LDME sets. Kinematic ranges are given in the legend boxes.}
\label{fig: lam-tung}
\end{figure}

Not all the invariants belong to the previous family. Indeed, one can exploit another relation that involves all polarization parameters in two frames and that, upon rotation around the $Y$-axis, reads
\begin{equation}
    (\lambda_{F'} - \nu_{F'}/2)^2 + 4 \mu_{F'}^2 = {(\lambda_{F} - \nu_{F}/2)^2 + 4 \mu_{F}^2 \over (1 + \rho)^2} \, .
\end{equation}

From this, one can construct an invariant quantity involving the polarization parameters squared, as first pointed out in
Ref.~\cite{Palestini:2010xu}. As an example, we recall
\begin{equation}
	\tilde \lambda' = {(\l - \nu/2)^2 + 4 \mu^2 \over (3 + \l)^2}
	\, ,
	\label{eq: invariant tilde lambda}
\end{equation}
as introduced in Ref.~\cite{Peng:2018tty}.

The study of rotational invariants has not only a theoretical interest, but it is relevant also from the experimental point of view, since their expected equality among different frames is an important check of experimental acceptances and systematics as shown, for instance, by the ATLAS Collaboration~\cite{ATLAS:2011aqv}.

For these reasons, we consider, as a case of study, one of these quantities at the kinematics explored by the EIC. In Fig.~\ref{fig: lam-tung} we show the theoretical estimates in the collinear framework, for the invariant ${\cal F}$, Eq.~({\ref{eq: invariant F1}}), as a function of $z$ (left panels) and $P_T$ (right panels). Once again we compute this quantity at two energies, $\sqrt s = 140$~GeV (upper panels) and $\sqrt s = 45$~GeV (lower panels) for different approaches and LDME sets.

From Fig.~\ref{fig: lam-tung} we clearly see that ${\cal F}$ is not equal to $1/2$, as expected from the Lam-Tung relation. Moreover, it is neither a constant, since its value depends on both $z$ and $P_T$ variables.
In principle, for some LDME sets a constant behavior could accidentally appear, but this would be limited to a specific kinematic region.

Another interesting remark is that, while the denominator of ${\cal F}$ is proportional to the unpolarized cross section, its numerator is controlled by the relative size of the $\lambda$ and $\nu$ parameters.
This can vary significantly,
depending on the frames and approaches adopted, as discussed in the  previous Section.

From this preliminary study we can conclude that, even if not easily accessible from the experimental point of view, these invariant quantities could represent an invaluable tool to learn on the $J/\psi$ polarization mechanism.

\section{Conclusions}
\label{sec: conclusions}
The study of quarkonium polarization, interesting by itself, is also a powerful tool to explore the still challenging issue of its formation mechanism within QCD. In this spirit, we have presented a phenomenological analysis of $J/\psi$ polarization in SIDIS at large $P_T$. More specifically, we have looked at the dilepton angular distribution in the $J/\psi\to \ell^+\ell^-$ decay in terms of the associated polarization parameters, that could be accessed at the future EIC. By exploiting the theoretical results of Ref.~\cite{DAlesio:2021yws}, we have computed the parameters, $\lambda$, $\mu$ and $\nu$, in different frames, trying to emphasize whether one can use these observables to discriminate among two well consolidated frameworks, still under investigation: the Color Singlet Model and the NRQCD approach. Moreover, for the latter we have employed three different LDME sets, based on different extractions and assumptions, highlighting their impact on quarkonium polarization estimates.

We have shown results both as a function of $z$ and $P_T$, adopting two quite different cm energies, for standard kinematics at the EIC, together with a detailed analysis in terms of parton and NRQCD wave contributions.

The main findings of our study can be summarized as follows: $i)$ concerning the $\lambda$ parameter, the large-$z$ region, both in the Gottfried-Jackson and the Helicity frame, turns out to be very promising, with the only caveat of possible contributions from (TMD) shape functions (even if expected to be reduced being $\lambda$ a ratio of helicity structure functions); similarly its $P_T$ distribution, at medium-large values, could be an ideal ground to disentangle the formation mechanisms, both at high and low energies. $ii)$ The $\mu$ parameter displays some interesting features when studied in the Gottfried-Jackson frame, namely: a clear separation among the estimates in different frameworks at medium-large $z$ or as a function of $P_T$ in the high-energy set-up; a different behavior with respect to the corresponding lower-energy estimates at medium-large $z$ or at moderate $P_T$. Moreover, in the Helicity frame at low energies one could extract important information by looking in the large $P_T$ region.
$iii)$ Similarly, for the $\nu$ parameter, relevant also in the context of the TMD framework, medium-large $P_T$ values in the Gottfried-Jackson frame are certainly worth to be explored.

Finally, we have discussed a selection of frame-independent (rotational invariant) polarization parameters, relevant not only from the theory point of view, but extremely useful as an important check of experimental acceptances and systematics. In particular, we have focused on the invariant ${\cal{F}}$, controlled by the relative weight of the $\lambda$ and $\nu$ parameters, that strongly depend on the frames and frameworks adopted.
As shown, this observable could clearly help in getting information on the $J/\psi$ formation mechanism, both at large $z$ (high- and low-energy set-ups) and as a function of $P_T$ (at large energy).

We can certainly conclude that a study of the dilepton angular distribution in $J/\psi$ decay in SIDIS at the EIC could be an invaluable tool to shed light on the $J/\psi$ polarization as well as on its formation mechanism.

\acknowledgments
We thank P.~Faccioli, T.~Stebel and R.~Venugopalan for clarifying some aspects concerning the rotational invariants. This project has received funding from the European Union’s Horizon 2020 research and innovation programme under grant agreement STRONG 2020—No 824093. U.D.~and C.P.~also acknowledge financial support by Fondazione di Sardegna under the project ``Proton
tomography at the LHC'', project number F72F20000220007 (University of Cagliari).

\bibliography{Bibliography}{}

\providecommand{\href}[2]{#2}\begingroup\raggedright\begin{thebibliography}{10}

\bibitem{E598:1974sol}
{\bfseries E598} Collaboration, J.~J. Aubert {\em et~al.}, ``{Experimental
  Observation of a Heavy Particle $J$},''
  \href{http://dx.doi.org/10.1103/PhysRevLett.33.1404}{{\em Phys. Rev. Lett.}
  {\bfseries 33} (1974) 1404--1406}.

\bibitem{SLAC-SP-017:1974ind}
{\bfseries SLAC-SP-017} Collaboration, J.~E. Augustin {\em et~al.},
  ``{Discovery of a Narrow Resonance in $e^+ e^-$ Annihilation},''
  \href{http://dx.doi.org/10.1103/PhysRevLett.33.1406}{{\em Phys. Rev. Lett.}
  {\bfseries 33} (1974) 1406--1408}.

\bibitem{Baier:1983va}
R.~Baier and R.~Ruckl, ``{Hadronic collisions: A quarkonium factory},''
  \href{http://dx.doi.org/10.1007/BF01572254}{{\em Z. Phys. C} {\bfseries 19}
  (1983) 251}.

\bibitem{Bodwin:1994jh}
G.~T. Bodwin, E.~Braaten, and G.~P. Lepage, ``{Rigorous QCD analysis of
  inclusive annihilation and production of heavy quarkonium},''
  \href{http://dx.doi.org/10.1103/PhysRevD.51.1125}{{\em Phys. Rev. D}
  {\bfseries 51} (1995) 1125--1171},
  \href{http://arxiv.org/abs/hep-ph/9407339}{{\ttfamily arXiv:hep-ph/9407339}}.
  [Erratum: Phys.Rev.D 55, 5853 (1997)].

\bibitem{Nayak:2006fm}
G.~C. Nayak, J.-W. Qiu, and G.~F. Sterman, ``{NRQCD factorization and the
  velocity dependence of NNLO poles in heavy quarkonium production},''
  \href{http://dx.doi.org/10.1103/PhysRevD.74.074007}{{\em Phys. Rev. D}
  {\bfseries 74} (2006) 074007},
  \href{http://arxiv.org/abs/hep-ph/0608066}{{\ttfamily arXiv:hep-ph/0608066}}.

\bibitem{Lepage:1992tx}
G.~P. Lepage, L.~Magnea, C.~Nakhleh, U.~Magnea, and K.~Hornbostel, ``{Improved
  nonrelativistic QCD for heavy-quark physics},''
  \href{http://dx.doi.org/10.1103/PhysRevD.46.4052}{{\em Phys. Rev. D}
  {\bfseries 46} (1992) 4052--4067},
  \href{http://arxiv.org/abs/hep-lat/9205007}{{\ttfamily
  arXiv:hep-lat/9205007}}.

\bibitem{Butenschoen:2010rq}
M.~Butenschoen and B.~A. Kniehl, ``{Reconciling $J/\psi$ production at HERA,
  RHIC, Tevatron, and LHC with NRQCD factorization at next-to-leading order},''
  \href{http://dx.doi.org/10.1103/PhysRevLett.106.022003}{{\em Phys. Rev.
  Lett.} {\bfseries 106} (2011) 022003},
  \href{http://arxiv.org/abs/1009.5662}{{\ttfamily arXiv:1009.5662 [hep-ph]}}.

\bibitem{Chao:2012iv}
K.-T. Chao, Y.-Q. Ma, H.-S. Shao, K.~Wang, and Y.-J. Zhang, ``{$J/\psi$
  Polarization at Hadron Colliders in Nonrelativistic QCD},''
  \href{http://dx.doi.org/10.1103/PhysRevLett.108.242004}{{\em Phys. Rev.
  Lett.} {\bfseries 108} (2012) 242004},
  \href{http://arxiv.org/abs/1201.2675}{{\ttfamily arXiv:1201.2675 [hep-ph]}}.

\bibitem{Sharma:2012dy}
R.~Sharma and I.~Vitev, ``{High transverse momentum quarkonium production and
  dissociation in heavy ion collisions},''
  \href{http://dx.doi.org/10.1103/PhysRevC.87.044905}{{\em Phys. Rev. C}
  {\bfseries 87} no.~4, (2013) 044905},
  \href{http://arxiv.org/abs/1203.0329}{{\ttfamily arXiv:1203.0329 [hep-ph]}}.

\bibitem{Bodwin:2014gia}
G.~T. Bodwin, H.~S. Chung, U.-R. Kim, and J.~Lee, ``{Fragmentation
  contributions to $J/\psi$ production at the Tevatron and the LHC},''
  \href{http://dx.doi.org/10.1103/PhysRevLett.113.022001}{{\em Phys. Rev.
  Lett.} {\bfseries 113} no.~2, (2014) 022001},
  \href{http://arxiv.org/abs/1403.3612}{{\ttfamily arXiv:1403.3612 [hep-ph]}}.

\bibitem{Zhang:2014ybe}
H.-F. Zhang, Z.~Sun, W.-L. Sang, and R.~Li, ``{Impact of $\eta_c$
  hadroproduction data on charmonium production and polarization within the
  NRQCD framework},''
  \href{http://dx.doi.org/10.1103/PhysRevLett.114.092006}{{\em Phys. Rev.
  Lett.} {\bfseries 114} no.~9, (2015) 092006},
  \href{http://arxiv.org/abs/1412.0508}{{\ttfamily arXiv:1412.0508 [hep-ph]}}.

\bibitem{Brambilla:2010cs}
N.~Brambilla {\em et~al.}, ``{Heavy Quarkonium: Progress, Puzzles, and
  Opportunities},''
  \href{http://dx.doi.org/10.1140/epjc/s10052-010-1534-9}{{\em Eur. Phys. J. C}
  {\bfseries 71} (2011) 1534}, \href{http://arxiv.org/abs/1010.5827}{{\ttfamily
  arXiv:1010.5827 [hep-ph]}}.

\bibitem{Andronic:2015wma}
A.~Andronic {\em et~al.}, ``{Heavy-flavour and quarkonium production in the LHC
  era: from proton\textendash{}proton to heavy-ion collisions},''
  \href{http://dx.doi.org/10.1140/epjc/s10052-015-3819-5}{{\em Eur. Phys. J. C}
  {\bfseries 76} no.~3, (2016) 107},
  \href{http://arxiv.org/abs/1506.03981}{{\ttfamily arXiv:1506.03981
  [nucl-ex]}}.

\bibitem{Lansberg:2019adr}
J.-P. Lansberg, ``{New observables in inclusive production of quarkonia},''
  \href{http://dx.doi.org/10.1016/j.physrep.2020.08.007}{{\em Phys. Rept.}
  {\bfseries 889} (2020) 1--106},
  \href{http://arxiv.org/abs/1903.09185}{{\ttfamily arXiv:1903.09185
  [hep-ph]}}.

\bibitem{Zhang:2019ecf}
H.-F. Zhang, W.-L. Sang, and Y.-P. Yan, ``{Statistical analysis of the
  azimuthal asymmetry in the $J/\psi$ leptoproduction in unpolarized $ep$
  collisions},'' \href{http://dx.doi.org/10.1007/JHEP10(2019)234}{{\em JHEP}
  {\bfseries 10} (2019) 234}, \href{http://arxiv.org/abs/1908.02521}{{\ttfamily
  arXiv:1908.02521 [hep-ph]}}.

\bibitem{Qiu:2020xum}
J.-W. Qiu, X.-P. Wang, and H.~Xing, ``{Exploring $J/\psi$ Production Mechanism
  at the Future Electron-Ion Collider},''
  \href{http://dx.doi.org/10.1088/0256-307X/38/4/041201}{{\em Chin. Phys.
  Lett.} {\bfseries 38} no.~4, (2021) 041201},
  \href{http://arxiv.org/abs/2005.10832}{{\ttfamily arXiv:2005.10832
  [hep-ph]}}.

\bibitem{Boer:2021ehu}
D.~Boer, C.~Pisano, and P.~Taels, ``{Extracting color octet NRQCD matrix
  elements from $J/\psi$ production at the EIC},''
  \href{http://dx.doi.org/10.1103/PhysRevD.103.074012}{{\em Phys. Rev. D}
  {\bfseries 103} no.~7, (2021) 074012},
  \href{http://arxiv.org/abs/2102.00003}{{\ttfamily arXiv:2102.00003
  [hep-ph]}}.

\bibitem{AbdulKhalek:2021gbh}
R.~Abdul~Khalek {\em et~al.}, ``{Science Requirements and Detector Concepts for
  the Electron-Ion Collider}: {EIC Yellow Report},''
  \href{http://dx.doi.org/10.1016/j.nuclphysa.2022.122447}{{\em Nucl. Phys. A}
  {\bfseries 1026} (2022) 122447},
  \href{http://arxiv.org/abs/2103.05419}{{\ttfamily arXiv:2103.05419
  [physics.ins-det]}}.

\bibitem{Accardi:2012qut}
A.~Accardi {\em et~al.}, ``{Electron Ion Collider: The next QCD frontier -
  Understanding the glue that binds us all},''
  \href{http://dx.doi.org/10.1140/epja/i2016-16268-9}{{\em Eur. Phys. J. A}
  {\bfseries 52} no.~9, (2016) 268},
  \href{http://arxiv.org/abs/1212.1701}{{\ttfamily arXiv:1212.1701 [nucl-ex]}}.

\bibitem{Boer:2011fh}
D.~Boer {\em et~al.}, ``{Gluons and the quark sea at high energies:
  Distributions, polarization, tomography},''
  \href{http://arxiv.org/abs/1108.1713}{{\ttfamily arXiv:1108.1713 [nucl-th]}}.

\bibitem{H1:2002xeb}
{\bfseries H1} Collaboration, C.~Adloff {\em et~al.}, ``{Inelastic
  leptoproduction of $J/\psi$ mesons at HERA},''
  \href{http://dx.doi.org/10.1007/s10052-002-1014-y}{{\em Eur. Phys. J. C}
  {\bfseries 25} (2002) 41--53},
  \href{http://arxiv.org/abs/hep-ex/0205065}{{\ttfamily arXiv:hep-ex/0205065}}.

\bibitem{Fleming:1997fq}
S.~Fleming and T.~Mehen, ``{Leptoproduction of $J/\psi$},''
  \href{http://dx.doi.org/10.1103/PhysRevD.57.1846}{{\em Phys. Rev. D}
  {\bfseries 57} (1998) 1846--1857},
  \href{http://arxiv.org/abs/hep-ph/9707365}{{\ttfamily arXiv:hep-ph/9707365}}.

\bibitem{Yuan:2000cn}
F.~Yuan and K.-T. Chao, ``{Polarized $J/\psi$ production in deep inelastic
  scattering at DESY HERA},''
  \href{http://dx.doi.org/10.1103/PhysRevD.63.034017}{{\em Phys. Rev. D}
  {\bfseries 63} (2001) 034017},
  \href{http://arxiv.org/abs/hep-ph/0008301}{{\ttfamily arXiv:hep-ph/0008301}}.
  [Erratum: Phys.Rev.D 66, 079902 (2002)].

\bibitem{Sun:2017wxk}
Z.~Sun and H.-F. Zhang, ``{QCD corrections to the color-singlet $J/\psi$
  production in deeply inelastic scattering at HERA},''
  \href{http://dx.doi.org/10.1103/PhysRevD.96.091502}{{\em Phys. Rev. D}
  {\bfseries 96} no.~9, (2017) 091502},
  \href{http://arxiv.org/abs/1705.05337}{{\ttfamily arXiv:1705.05337
  [hep-ph]}}.

\bibitem{DAlesio:2021yws}
U.~D'Alesio, L.~Maxia, F.~Murgia, C.~Pisano, and S.~Rajesh, ``{$J/\psi$
  polarization in semi-inclusive DIS at low and high transverse momentum},''
  \href{http://dx.doi.org/10.1007/JHEP03(2022)037}{{\em JHEP} {\bfseries 03}
  (2022) 037}, \href{http://arxiv.org/abs/2110.07529}{{\ttfamily
  arXiv:2110.07529 [hep-ph]}}.

\bibitem{Lam:1978pu}
C.~S. Lam and W.-K. Tung, ``{Systematic approach to inclusive lepton pair
  production in hadronic collisions},''
  \href{http://dx.doi.org/10.1103/PhysRevD.18.2447}{{\em Phys. Rev. D}
  {\bfseries 18} (1978) 2447}.

\bibitem{Beneke:1998re}
M.~Beneke, M.~Kramer, and M.~Vanttinen, ``{Inelastic photoproduction of
  polarized $J/\psi$},'' \href{http://dx.doi.org/10.1103/PhysRevD.57.4258}{{\em
  Phys. Rev. D} {\bfseries 57} (1998) 4258--4274},
  \href{http://arxiv.org/abs/hep-ph/9709376}{{\ttfamily arXiv:hep-ph/9709376}}.

\bibitem{Gong:2012ug}
B.~Gong, L.-P. Wan, J.-X. Wang, and H.-F. Zhang, ``{Polarization for Prompt
  $J/\psi$ and $\psi(2s)$ Production at the Tevatron and LHC},''
  \href{http://dx.doi.org/10.1103/PhysRevLett.110.042002}{{\em Phys. Rev.
  Lett.} {\bfseries 110} no.~4, (2013) 042002},
  \href{http://arxiv.org/abs/1205.6682}{{\ttfamily arXiv:1205.6682 [hep-ph]}}.

\bibitem{Butenschoen:2011yh}
M.~Butenschoen and B.~A. Kniehl, ``{World data of $J/\psi$ production
  consolidate NRQCD factorization at next-to-leading order},''
  \href{http://dx.doi.org/10.1103/PhysRevD.84.051501}{{\em Phys. Rev. D}
  {\bfseries 84} (2011) 051501},
  \href{http://arxiv.org/abs/1105.0820}{{\ttfamily arXiv:1105.0820 [hep-ph]}}.

\bibitem{Pumplin:2002vw}
J.~Pumplin, D.~R. Stump, J.~Huston, H.~L. Lai, P.~M. Nadolsky, and W.~K. Tung,
  ``{New generation of parton distributions with uncertainties from global QCD
  analysis},'' \href{http://dx.doi.org/10.1088/1126-6708/2002/07/012}{{\em
  JHEP} {\bfseries 07} (2002) 012},
  \href{http://arxiv.org/abs/hep-ph/0201195}{{\ttfamily arXiv:hep-ph/0201195}}.

\bibitem{Beneke:1999gq}
M.~Beneke, G.~A. Schuler, and S.~Wolf, ``{Quarkonium momentum distributions in
  photoproduction and $B$ decay},''
  \href{http://dx.doi.org/10.1103/PhysRevD.62.034004}{{\em Phys. Rev. D}
  {\bfseries 62} (2000) 034004},
  \href{http://arxiv.org/abs/hep-ph/0001062}{{\ttfamily arXiv:hep-ph/0001062}}.

\bibitem{Beneke:1997qw}
M.~Beneke, I.~Z. Rothstein, and M.~B. Wise, ``{Kinematic enhancement of
  non-perturbative corrections to quarkonium production},''
  \href{http://dx.doi.org/10.1016/S0370-2693(97)00832-0}{{\em Phys. Lett. B}
  {\bfseries 408} (1997) 373--380},
  \href{http://arxiv.org/abs/hep-ph/9705286}{{\ttfamily arXiv:hep-ph/9705286}}.

\bibitem{Echevarria:2019ynx}
M.~G. Echevarria, ``{Proper TMD factorization for quarkonia production:
  $pp\to\eta_{c,b}$ as a study case},''
  \href{http://dx.doi.org/10.1007/JHEP10(2019)144}{{\em JHEP} {\bfseries 10}
  (2019) 144}, \href{http://arxiv.org/abs/1907.06494}{{\ttfamily
  arXiv:1907.06494 [hep-ph]}}.

\bibitem{Fleming:2019pzj}
S.~Fleming, Y.~Makris, and T.~Mehen, ``{An effective field theory approach to
  quarkonium at small transverse momentum},''
  \href{http://dx.doi.org/10.1007/JHEP04(2020)122}{{\em JHEP} {\bfseries 04}
  (2020) 122}, \href{http://arxiv.org/abs/1910.03586}{{\ttfamily
  arXiv:1910.03586 [hep-ph]}}.

\bibitem{Boer:2020bbd}
D.~Boer, U.~D'Alesio, F.~Murgia, C.~Pisano, and P.~Taels, ``{$J/\psi$ meson
  production in SIDIS: matching high and low transverse momentum},''
  \href{http://dx.doi.org/10.1007/JHEP09(2020)040}{{\em JHEP} {\bfseries 09}
  (2020) 040}, \href{http://arxiv.org/abs/2004.06740}{{\ttfamily
  arXiv:2004.06740 [hep-ph]}}.

\bibitem{Faccioli:2010ji}
P.~Faccioli, C.~Lourenco, and J.~Seixas, ``{New approach to quarkonium
  polarization studies},''
  \href{http://dx.doi.org/10.1103/PhysRevD.81.111502}{{\em Phys. Rev. D}
  {\bfseries 81} (2010) 111502},
  \href{http://arxiv.org/abs/1005.2855}{{\ttfamily arXiv:1005.2855 [hep-ph]}}.

\bibitem{Faccioli:2010ej}
P.~Faccioli, C.~Lourenco, and J.~Seixas, ``{Rotation-invariant relations in
  vector meson decays into fermion pairs},''
  \href{http://dx.doi.org/10.1103/PhysRevLett.105.061601}{{\em Phys. Rev.
  Lett.} {\bfseries 105} (2010) 061601},
  \href{http://arxiv.org/abs/1005.2601}{{\ttfamily arXiv:1005.2601 [hep-ph]}}.

\bibitem{Faccioli:2010kd}
P.~Faccioli, C.~Lourenco, J.~Seixas, and H.~K. Wohri, ``{Towards the
  experimental clarification of quarkonium polarization},''
  \href{http://dx.doi.org/10.1140/epjc/s10052-010-1420-5}{{\em Eur. Phys. J. C}
  {\bfseries 69} (2010) 657--673},
  \href{http://arxiv.org/abs/1006.2738}{{\ttfamily arXiv:1006.2738 [hep-ph]}}.

\bibitem{Faccioli:2011zzb}
P.~Faccioli, C.~Lourenco, J.~Seixas, and H.~K. Wohri, ``{Quarkonium
  polarization in $p p$ and $p$-nucleus collisions},''
  \href{http://dx.doi.org/10.1016/j.nuclphysa.2011.02.027}{{\em Nucl. Phys. A}
  {\bfseries 855} (2011) 116--124}.

\bibitem{Faccioli:2011zzc}
P.~Faccioli, C.~Lourenco, J.~Seixas, and H.~K. Wohri, ``{Quarkonium
  polarization measurements},''
  \href{http://dx.doi.org/10.1016/j.nuclphysbps.2011.03.065}{{\em Nucl. Phys. B
  Proc. Suppl.} {\bfseries 214} (2011) 97--102}.

\bibitem{Peng:2018tty}
J.-C. Peng, D.~Boer, W.-C. Chang, R.~E. McClellan, and O.~Teryaev, ``{On the
  rotational invariance and non-invariance of lepton angular distributions in
  Drell\textendash{}Yan and quarkonium production},''
  \href{http://dx.doi.org/10.1016/j.physletb.2018.11.061}{{\em Phys. Lett. B}
  {\bfseries 789} (2019) 356--359},
  \href{http://arxiv.org/abs/1808.04398}{{\ttfamily arXiv:1808.04398
  [hep-ph]}}.

\bibitem{Ma:2018qvc}
Y.-Q. Ma, T.~Stebel, and R.~Venugopalan, ``{$J/\psi$ polarization in the
  CGC+NRQCD approach},'' \href{http://dx.doi.org/10.1007/JHEP12(2018)057}{{\em
  JHEP} {\bfseries 12} (2018) 057},
  \href{http://arxiv.org/abs/1809.03573}{{\ttfamily arXiv:1809.03573
  [hep-ph]}}.

\bibitem{ALICE:2018crw}
{\bfseries ALICE} Collaboration, S.~Acharya {\em et~al.}, ``{Measurement of the
  inclusive $J/\psi $ polarization at forward rapidity in $pp$ collisions at
  $\sqrt{s} = 8$ TeV},''
  \href{http://dx.doi.org/10.1140/epjc/s10052-018-6027-2}{{\em Eur. Phys. J. C}
  {\bfseries 78} no.~7, (2018) 562},
  \href{http://arxiv.org/abs/1805.04374}{{\ttfamily arXiv:1805.04374
  [hep-ex]}}.

\bibitem{Faccioli:2011pn}
P.~Faccioli, C.~Lourenco, J.~Seixas, and H.~K. Wohri, ``{Model-independent
  constraints on the shape parameters of dilepton angular distributions},''
  \href{http://dx.doi.org/10.1103/PhysRevD.83.056008}{{\em Phys. Rev. D}
  {\bfseries 83} (2011) 056008},
  \href{http://arxiv.org/abs/1102.3946}{{\ttfamily arXiv:1102.3946 [hep-ph]}}.

\bibitem{Palestini:2010xu}
S.~Palestini, ``{Angular distribution and rotations of frame in vector meson
  decays into lepton pairs},''
  \href{http://dx.doi.org/10.1103/PhysRevD.83.031503}{{\em Phys. Rev. D}
  {\bfseries 83} (2011) 031503},
  \href{http://arxiv.org/abs/1012.2485}{{\ttfamily arXiv:1012.2485 [hep-ph]}}.

\bibitem{ATLAS:2011aqv}
{\bfseries ATLAS} Collaboration, G.~Aad {\em et~al.}, ``{Measurement of the
  differential cross-sections of inclusive, prompt and non-prompt $J/\psi$
  production in proton-proton collisions at $\sqrt{s}=7$ TeV},''
  \href{http://dx.doi.org/10.1016/j.nuclphysb.2011.05.015}{{\em Nucl. Phys. B}
  {\bfseries 850} (2011) 387--444},
  \href{http://arxiv.org/abs/1104.3038}{{\ttfamily arXiv:1104.3038 [hep-ex]}}.

\end{thebibliography}\endgroup
\bibliographystyle{utphys}

\end{document}